\documentstyle[11pt,fleqn]{article}    
\title{Macroscopic Estimate of the Average Speed \\ for New Particles Created in Collision}
\author{A. Kwang-Hua Chu }
\date{P.O. Box 39, Tou-Di-Ban, Road XiHong, Urumqi 830000}
\begin{document}           
\oddsidemargin+2.5mm
\doublerulesep=6mm      
\baselineskip=8mm
\maketitle             
\begin{abstract}
\noindent
A {\it heuristic} approach is proposed to estimate the average  speed of particles
during binary encounters by using the macroscopic variables with their
extended gradient-type
which are the fundamental independent variables in {\it extended
thermodynamics} theory. We also address the missing contribution
(say, due to creation of new particles : {\it Acoustons}) in conventional Bremsstrahlung.
\newline

\noindent
{\bf Key Words :} \hspace*{3mm} Dynamic Casimir effect; Bremsstrahlung;
stationary conservation form; hard-sphere.

\end{abstract}
\bibliographystyle{plain}
\section{Introduction}
The rough approximations to the velocity of particles for a specific distribution
have been a fundamental issue in the kinetic theory of gas \cite{Jeans:Gas}.
Normally the molecular speed varies with the molecular weight and
absolute temperature if we only consider the translation part of the
kinetic energy \cite{Chapman:Cowling} for a gas in a uniform steady
state. \newline
We shall discuss the approximate estimate of average particle speed (in 1-D.
sense, $c_x$) from the stationary equations of wave-breaking-like conservation
laws [3-4]. The {\it flow} is assumed to be
uniformly bounded and avoid the vacuum state. Since a conservation law
is an integral relation, it may be satisfied by functions which are not
differentiable (like the discrete particle- or molecule-based flow using
Boltzmann approach for {\it dilute} gas), not even continuous, merely
measurable and bounded. \newline
We noticed that steady shocks can occur in an ideal case [5-6]
or in a microscopic way [7-9]. In this short paper, we will investigate this kind of
1D flow in a heuristic way. The flow field (if in terms of the flow velocity)
depends on the pressure gradient and density gradient only.
\section{Formulation}
\subsection{Stationary Weak Shock}
Starting from the integral form of the balanced equations for the one-dimensional flow
allowing discontinuity in x-direction velocity $u$ (cf. Fig. 1):
\begin{equation}
 \frac{d}{dt}\int_{x_l}^{x_r} f dx +[g]_{x_l}^{x_r} =0,
\end{equation}
here, [$\,$] relates to the jump \cite{Whitham:Stoss}; $f$, $g$ can be the density
and flux of mass, or the density and flux of momentum, and we neglect the
source-term effects, e.g., body force in the momentum-balance analogy. We assume
that $u$ has continuous first derivatives and $f$, $g$ are functions of $x$,
$t$, $u$. Thus, together with jump condition \cite{Whitham:Stoss} and entropy
condition for a weak solution [3,4], (1) becomes
\begin{equation}
 \frac{\partial f(x,t,u)}{\partial t}+\frac{\partial g(x,t,u)}{\partial
 x} =0.
\end{equation}
Let $f$ be the density and flux of mass, $g$ be the flux of mass and
momentum, then we have
\begin{equation}
 \frac{\partial \rho}{\partial t}+\frac{\partial(\rho\,u)}{\partial x} =0,
\end{equation}
\begin{equation}
 \frac{\partial (\rho\,u)}{\partial t}+\frac{\partial(\rho\,u^2+p)}{\partial x}
 =0.
\end{equation}
Here,
\begin{displaymath}
p=-\frac{1}{3}p_{ii}=\frac{1}{3} \int^{\infty}_{-\infty} m c_i^2
F d{\bf c} ,
\end{displaymath}
$F$ is the velocity distribution function of molecules from the kinetic
theory of gases, $m$ is the mass of a molecule, ${\bf c}$ (or $c_i$) measures the
difference of the molecular velocity from the mean value or macroscopic
velocity $\int^{\infty}_{-\infty}$ $m {\bf v} f d{\bf v}$; ${\bf v}$ is the
absolute molecular velocity. \newline
The stationary solution $u$ from Eqns. (3) and (4) then is
\begin{equation}
 u=(\frac{\partial p/\partial x}{\partial \rho/\partial x})^{1/2} .
\end{equation}
For the cases of weak shocks, Eqn. (5) tends to the characteristic
velocity
\begin{equation}
 U=(\frac{\partial p}{\partial \rho})^{1/2},
\end{equation}
in the limit as the shock strength approaches zero, which is just the
generalization of 1D sound speed. Up to now, the internal energy $e$ or
the enthalpy $h=e+{p}/{\rho}$, which for the ideal gas is a function of
temperature alone, are still not specified yet. Besides, we neglect the
viscous and heat-conducting effects in general. Thus, this kind of
stationary shock can only exist either in discrete sense [11-12] or in microscopic way (e.g., induced by molecular
collisions) [9,13]. Considering the
time scale of collisions, e.g., the mean collision time , since it is much
shorter than the relaxation time, so, once we neglect the high-frequency
behavior or relaxation effects and only take the low-frequency limit into
account, then the {\it stationary shock} concept is valid.  \newline
\subsection{Application to Bremsstrahlung
and New Particles Creation}
The conservation equations obtained after we impose {\it weak} formulations from
the integral forms which are similar to the treatments of weak-shock
problems \cite{Lax:SIAM} are constructed from the collision diagram as shown
in Fig. 1 with respect to the axes $x$ and $y$. The one-dimensional velocity $c$
(in average) for particles during a binary encounter is in the x-direction
for simplicity.
The stationary equations, if we neglect the time-dependent effects, are
\begin{equation}
 (\rho c)_x =0,   \hspace*{6mm} (\rho c^2 + p)_x =0 .
\end{equation}
$c$ has been spatially and locally homogenized \cite{Allaire:Homogenization}
with $c$ $\in C^1$. Thus, similar to the derivation of $u$ (cf. the equation (5))
we can get the average estimate of $c$ as
\begin{displaymath}
 c=(\frac{p_x}{\rho_x})^{1/2} .
\end{displaymath}
Note that this velocity could be linked to the sound speed by the {\it extended
thermodynamics} theory [15-16] and might be related to the neglected (energy) contributions
during {\it Bremsstrahlung} (or collisions of particles) since it is rather weak (however, it should not be neglected
considering the strong and weak interactions of particles) compared to the photon
emission or others! To be precise, the approach we used above could be applied to the
dynamic Casimir effect and the more interesting manifestation of the dynamic behavior is the creation of particles from vacuum by a moving boundary (here, the moving boundary
is related to the non-flat shape) [17-19]! The vacuum fluctuations, according to the Casimir
effect, can generate the pressure field and thus the acoustic field as mentioned above!
The effect of creation of particles from vacuum by nonstationary electric and
gravitational fields is well known (see, e.g., [20,21]). As noted above, boundary conditions are idealizations of
concentrated external fields. It is not surprising, then, that moving boundaries act
in the same
way as a nonstationary external field. As to the possibility of experimental
observation of the photons created by the moving mirrors. Additional
factors such as imperfectness of the boundary mirrors, back reaction of the
radiated photons upon the mirror, etc., could be traced in [22].
\section{Results \& Discussions}
From Fig. 1 we know that, if we transform the coordinate system into the
one based on the mass-center of these two colliding particles or molecules,
as the particles are assumed to be hard-sphere ones and have equal mass,
so, the mass-center (located at the contact point or the cross of $x$ \&
$y$-axis) will move with the speed ($c_{av}$) of total momentum
divided by the total mass \cite{Reif:StPhys}. The collisions are assumed to be elastic. In
fact, $c_{av}$ is equivalent to the speed of one-dimensional shock front. The energy
associated to this velocity is rather weak (in intermediate regime) and thus
was neglected in conventional
Bremsstrahlung (emission of photons or other radiations) but, as mentioned above,
it should be considered in the weak and strong interactions. The remaining question
is how to detect this kind of energy in the test section for creation of new
particles (say, if we termed these particles as {\it Acoustons}) subjected to collisions.  \newline

\setlength{\unitlength}{1.00mm}   
\begin{picture}(180,65)(10,-30)
\thinlines
\put(77,-3){\circle{11.0}}
\put(82,7){\circle{11.0}}
\thicklines
\put(77,-3){\vector(1,-1){15}}
\put(77,-3){\vector(1,2){16}}
\put(77,-3){\vector(2,1){15}}
\put(70,18){\makebox(0,0)[bl]{\large {\bf $p'_1$}}}
\put(61,0){\makebox(0,0)[bl]{\large {\bf $p_1$}}}
\put(82,7){\vector(-3,2){15}}
\put(82,7){\vector(-2,-1){15}}
\put(94,1){\makebox(0,0)[bl]{\large {\bf $p$}}}
\put(82,-15){\makebox(0,0)[bl]{\large {\bf $p'$}}}
\thinlines
\put(80,2){\line(2,-1){30}}
\put(80,2){\line(-2,1){30}}
\put(81,2){\makebox(0,0)[bl]{\small {\bf $G$}}}
\put(80,2){\circle*{1.0}}
\put(112,-14){\makebox(0,0)[bl]{\large {\bf $y$}}}
\put(46,18){\makebox(0,0)[bl]{\large {\bf $y$}}}
\put(70,-3){\line(4,0){30}}
\put(82,7){\line(2,0){15}}
\put(95,27){\makebox(0,0)[bl]{\large {\bf $x$}}}
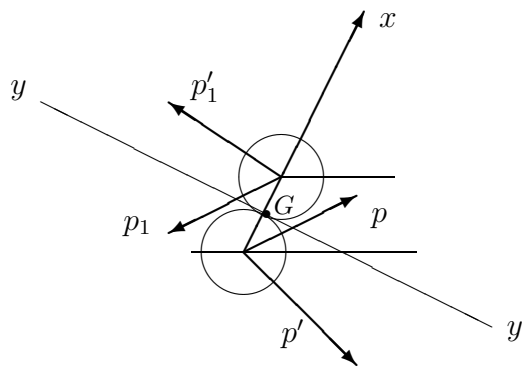
\put(30,-32){\makebox(0,0)[bl]{\large {Fig. 1\hspace*{2mm} A {\it head-on}
collision; after homogenization.}}}
\end{picture}
\vspace{4mm}

\subsubsection*{Acknowledgements} This work was the extension part of the
author's PhD thesis (dated 1997-Dec.) [16].



\end{document}